\documentclass[letter,twocolumn]{jpsj3}
\usepackage{txfonts}
\usepackage{bm}
\usepackage{braket}
\usepackage{bxpapersize}

\title{ Double-$Q$ Chiral Stripe in the $d$-$p$ Model with Strong Spin-Charge Coupling}
\author{Ryota Yambe and Satoru Hayami\thanks{Present address: Department of Applied Physics, The University of Tokyo, Tokyo 113-8656, Japan}}
\inst{Department of Physics, Hokkaido University, Sapporo 060-0810, Japan}

\abst{
We investigate the stability of multiple-$Q$ spiral states in $d$-$p$ electron systems with the strong spin-charge coupling.
By using variational calculations on a square lattice, we find that the double-$Q$ state with the scalar chirality density wave, which has been studied in the weak spin-charge coupling regime, becomes the ground state even in the strong spin-charge coupling regime by considering the effect of the $d$-$p$ hybridization. 
We also show that the regions where the double-$Q$ state is stabilized are widely extended for introducing the antiferromagnetic superexchange interaction between nearest-neighbor localized spins.  
}

\begin{document}
\maketitle

In condensed matter physics, noncoplanar spin textures have attracted great interest, as they give rise to unusual electronic states and transport phenomena in itinerant electrons through the spin-charge coupling.
The itinerant electrons acquire a gauge flux through the spin Berry phase in noncoplanar itinerant magnets, which results in a topological Hall effect and chiral spin liquids~\cite{PhysRevB.45.13544, PhysRevLett.83.3737, PhysRevB.62.R6065, PhysRevLett.87.116801, taguchi2001spin, PhysRevLett.101.156402,PhysRevLett.105.266405,batista2016frustration}.

The noncoplanar spin textures can often be characterized by a superposition of different ordering wave vectors, which are the so-called multiple-$Q$ states.
There are several mechanisms for stabilizing the multiple-$Q$ states, such as the Dzyaloshinskii-Moriya interaction~\cite{DZYALOSHINSKY1958241, PhysRev.120.91}
in chiral and polar magnets 
~\cite{roessler2006spontaneous, PhysRevLett.96.207202, PhysRevB.80.054416,PhysRevB.92.134405} and competitive exchange interactions in frustrated magnets.~\cite{PhysRevLett.108.017206, leonov2015multiply, PhysRevB.93.064430, PhysRevB.93.184413}.
Meanwhile, the spin-charge coupling in itinerant magnets provides another way to realize multiple-$Q$ states owing to the emergence of effective higher-order multiple-spin interactions~\cite{PhysRevLett.108.096401, ozawa2016vortex, PhysRevB.95.224424} besides the Ruderman-Kittel-Kasuya-Yosida (RKKY) interaction~\cite{PhysRev.96.99, 10.1143/PTP.16.45, PhysRev.106.893}.  
For example, various types of multiple-$Q$ states including magnetic skyrmion and vortex crystals have been explored in the Hubbard model~\cite{PhysRevLett.101.156402} and Kondo lattice model even without the spin-orbit coupling when the spin-charge coupling is small enough compared to the bandwidth of  itinerant electrons~\cite{akagi2010spin,barros2013efficient,PhysRevB.90.060402, ozawa2016vortex, PhysRevLett.118.147205, PhysRevB.99.094420}.
These theoretical investigations are important for clarifying the origin of multiple-$Q$ magnetic orderings found in various compounds, such as CeAuSb~\cite{PhysRevLett.120.097201}, Y$_3$Co$_8$Sn$_4$~\cite{takagi2018multiple}, Gd$_2$SiPd$_3$~\cite{kurumaji2019skyrmion}, and Gd$_3$Ru$_4$Al$_{12}$.~\cite{hirschberger2018skyrmion}

\begin{figure}[t]
\begin{center}
\includegraphics[width=8cm]{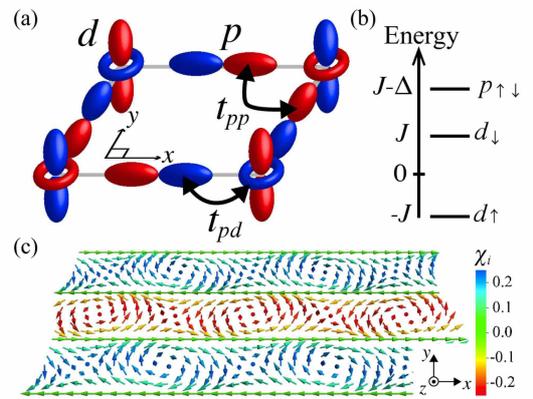}
\end{center}
\caption{(Color online). 
 (a) Schematic picture of the model in Eq.~(\ref{H}).
 $d$ orbitals form a square lattice and $\sigma$-bond $p$ orbitals are located at the bond center between the $d$ orbitals, whose colors represent the phase of the wave functions. 
 The hoppings $t_{pd}$ between the $d$ and $p$ orbitals and $t_{pp}$  between the $p$ orbitals are also shown.
(b) Schematic picture of the atomic energy levels of the $d$ and $p$ orbitals for the negative $\Delta$ in the FM state.
The spin degeneracy of the $d$ orbital split due to the Hund's-rule coupling through the other four $d$ orbitals.
 (c)  Schematic spin configuration of the double-$Q$ chiral stripe state at $\bm{Q}_1=(\pi /6,0)$, $\bm{Q}_2=(0,\pi /6)$, and $b=1$ in Eq.~(\ref{2Q}). 
The arrows represent the spin, and their colors represent the scalar spin chirality $\chi_i=\bm{S}_i \cdot ( \bm{S}_{i+x} \times \bm{S}_{i+y}) $ at site $i$.
} 
\label{fig1}
\end{figure} 

In the present study, we explore the possibility of stabilizing the multiple-$Q$ states by focusing on the strong spin-charge coupling compared to the bandwidth of itinerant electrons.
In this case, the system usually exhibits a ferromagnetic (FM) order for general electron filling by the double-exchange interaction between localized spins through the kinetic motion of itinerant electrons~\cite{PhysRev.82.403, PhysRev.100.675, PhysRev.118.141}. 
On the other hand, M. Mostovoy pointed out that the FM state is replaced with the single-$Q$ spiral state by taking into account the effect of the $d$-$p$ hybridization
in the double-exchange limit with the 3$d$ cubic perovskite SrFeO$_3$ in mind~\cite{mostovoy2005helicoidal,mostovoy2005helicoidal_2}.
However, recent experiments indicate that SrFeO$_3$ exhibits a plethora of multiple-$Q$ states in addition to the single-$Q$ spiral state in a wide range of temperatures and magnetic fields~\cite{PhysRevB.84.054427, ishiwata2018emergent, PhysRevMaterials.3.084404}. 
Thus, exploring the multiple-$Q$ states in $d$-$p$ electron systems with the strong spin-charge coupling is important.

In this Letter, we study the stability of multiple-$Q$ states when the effect of the $d$-$p$ hybridization and strong spin-charge coupling are taken into account.
By using variational calculations on a square lattice, we find that the energy of the double-$Q$ state is lower than that of both the single-$Q$ spiral and the FM states owing to the interplay between $p$-electron hopping and the negative charge transfer energy. 
The obtained double-$Q$ state accompanies the chirality density wave, whose spin pattern is similar to that in the Kondo lattice model with the weak spin-charge coupling~\cite{ozawa2016vortex}.
Moreover, we show that the effective exchange interaction that favors the double-$Q$ state is related to the effective $d$-$p$ hybridization rather than the direct spin-charge coupling.
We also show that the antiferromagnetic (AFM) superexchange interaction stabilizes the double-$Q$ state when the ordering vectors lie in the nearest-neighbor bond directions.

Let us consider the $d$-$p$ model with the double-exchange interaction on a square lattice.
We suppose that magnetic ions with $d$ electrons form a square network and oxygens with $p$ electrons are located at the bond center between nearest-neighbor magnetic ions
[Fig.~\ref{fig1}(a)].
The five $d$ orbitals split into three single levels, $d_{3z^2-r^2}$, $d_{x^2-y^2}$, and $d_{xy}$, and a doubly-degenerated level, ($d_{yz}$, $d_{zx}$) under the tetragonal crystalline electric field. 
By taking appropriate crystal-field parameters, we consider an itinerant $d_{3z^2-r^2}$ orbital coupled with four other localized $d$ orbitals through the Hund's-rule coupling~\cite{comment3} and the $p$ orbital along the $\sigma$ bond, which is similar to the model in SrFeO$_3$~\cite{mostovoy2005helicoidal, mostovoy2005helicoidal_2}.  
The Hamiltonian is given by
\begin{align}
\mathcal{H}&=t_{pd}\sum_{i,c,\sigma} \left(d^{\dagger}_{i\sigma}P_{ic\sigma}^{}+P^{\dagger}_{ic\sigma}d_{i\sigma}^{}\right)-t_{pp}\sum_{i,c\neq c',\sigma} P^{\dagger}_{ic\sigma}P_{ic'\sigma}^{} \nonumber\\
&\quad +(J-\Delta)\sum_{i,c,\sigma} p^{\dagger}_{i+\frac{c}{2}\sigma}p_{i+\frac{c}{2}\sigma}^{} -J\sum_{i,\sigma,\sigma'} d^{\dagger}_{i\sigma} \bm{\sigma}_{\sigma\sigma'}^{}d_{i\sigma'}^{} \cdot \bm{S}_i,  
 \label{H}
\end{align}       
where $d^{\dagger}_{i\sigma}(d_{i\sigma})$ and $p^{\dagger}_{i\sigma}(p_{i\sigma})$ are creation (annihilation) operators of $d_{3z^2-r^2}$ and $p$ electrons at site $i$ and spin $\sigma$, respectively, and $P_{ic\sigma}=p_{i+c/2\sigma}+p_{i-c/2\sigma}$ for $c=x,y$. 
The first and second terms represent electron hoppings between the $d$ and $p$ orbitals $t_{pd}$ and between the $p$ orbitals $t_{pp}$, where we adopt the Slater-Koster parameters by $t_{pd}=(0.5,0.5)(pd\sigma)$ and $t_{pp}=0.5\{(pp\sigma)-(pp\pi)\}$.
The third term is the $p$ energy level including the charge transfer energy $\Delta$, as shown in Fig.~\ref{fig1}(b). 
The fourth term is the strong Hund's-rule (spin-charge) coupling $J$ between $d_{3z^2-r^2}$ electron spins and localized spins $\bm{S}_i$ originating from other $d$ electron spins. 
We regard $\bm{S}_i$ as the classical spin with a length $|\bm{S}_i|=1$ for simplicity.
$\bm{\sigma}=(\sigma_x,\sigma_y,\sigma_z)$ is the Pauli matrix.
Hereafter, we take $(pd\sigma)=1$ and $a=1$ (lattice constant) as the energy and length units, respectively.
We adopt a strong spin-charge coupling $J=100$ compared to the bandwidth.

In the model in Eq.~(\ref{H}), the FM state becomes the ground state when the $p$ level is sufficiently far away from the Fermi level, as the model reduces to the double-exchange model. 
Meanwhile, the single-$Q$ spiral state can become the ground state when the $p$ level is relatively close to the Fermi level~\cite{mostovoy2005helicoidal,mostovoy2005helicoidal_2}. 
We here concentrate on the electron density where the continuous phase transition occurs from the FM state to the single-$Q$ spiral state by changing the model parameters, as discussed in Refs.~\citen{mostovoy2005helicoidal,mostovoy2005helicoidal_2}.
We find several electron densities that satisfy such a condition~\cite{comment2}, and we choose the electron density at $\sum_{i\sigma}\langle d_{i\sigma}^{\dagger}d_{i\sigma}^{} +\sum_c p_{i+
c/2\sigma}^{\dagger}p_{i+c/2\sigma}^{}\rangle=11/6$, where $\langle \cdots \rangle$ means the expectation value per unit cell.

To investigate the instability toward the multiple-$Q$ states in the ground state, 
we use the variational calculations for different magnetic patterns. 
For each magnetic state, we compute the internal energy $E$ at zero temperature by diagonalizing the Hamiltonian under the periodic boundary condition and determine the lowest-energy state.
We assume the FM, staggered-type AFM, single-$Q$ spiral, and double-$Q$ states.
The real-space spin configuration of the first three magnetic states is represented by
\begin{align}
\bm{S}^{1Q}_i=(0,\cos\bm{Q}_1\cdot\bm{r}_i,\sin\bm{Q}_1\cdot\bm{r}_i)^T, 
\label{1Q}
\end{align}
where $\bm{Q}_1$ is the ordering wave vector, $\bm{r}_i$ is the position vector of the site $i$, and $T$ is the transpose of the vector.
$\bm{Q}_1$ represents the variational parameters, where we consider three-types of ordering vectors: $\bm{Q}_1=(\phi,0)$ (denoted as A-type), $\bm{Q}_1=(\pi,\phi)$ (denoted as B-type), and $\bm{Q}_1=(\phi,\phi)$ (denoted as G-type). 
The spiral pitch $\phi$ $(0 \le \phi \le\pi)$ is given by $2\pi n/N_Q$ under the periodic boundary conditions, where $n$ is a non-negative integer and ($N_Q/2+1$) is the number of spiral pitches.
The spin configuration in Eq.~(\ref{1Q}) represents the FM and staggered-type AFM by setting $\phi=0$ and $\phi=\pi$, respectively, i.e., A(G)-type AFM is characterized by $\bm{Q}_1=(\pi, 0)$ [$\bm{Q}_1=(\pi, \pi)$].

The spin ansatz of the double-$Q$ state is given by~\cite{ozawa2016vortex}
\begin{align}
\bm{S}^{2Q}_i&=
\begin{pmatrix}
b\sin\bm{Q}_2\cdot\bm{r}_i\\
\sqrt{1-b^2+b^2\cos^2\bm{Q}_2\cdot\bm{r}_i}\cos\bm{Q}_1\cdot\bm{r}_i\\
\sqrt{1-b^2+b^2\cos^2\bm{Q}_2\cdot\bm{r}_i}\sin\bm{Q}_1\cdot\bm{r}_i
\end{pmatrix},
\label{2Q}
\end{align}
where the ordering vector $\bm{Q}_2$ is obtained by rotating $\bm{Q}_1$ by $ \pi/2$ around the $z$ axis, and $b$ $(0 < b \leq 1)$ represents the amplitude of the $\bm{Q}_2$ component.
Note that the spin configuration in Eq.~(\ref{2Q}) corresponds to that in the single-$Q$ spiral state at $b=0$. 
This double-$Q$ spin ansatz has been discussed in the Kondo lattice model with the small spin-charge coupling~\cite{ozawa2016vortex,PhysRevB.95.224424}. 
$\bm{Q}_1$ and $b$ are the variational parameters.

The real-space spin configuration of the double-$Q$ state is shown in Fig.~\ref{fig1}(c). 
The spin configuration is noncoplanar with different intensities at $\bm{Q}_1$ and $\bm{Q}_2$, and is characterized by a vortex-antivortex crystal 
without a uniform magnetization. 
Reflecting the noncoplanarity, this state possesses the scalar spin chirality degree of freedom $\chi_i=\bm{S}_i \cdot ( \bm{S}_{i+x} \times \bm{S}_{i+y}) $ at site $i$, as represented by the color plot in Fig.~\ref{fig1}(c).
In fact, the scalar chirality forms stripes in the $\bm{Q}_2$ direction, although its net component vanishes owing to the cancellation of the contribution from the vortices and antivortices~\cite{ozawa2016vortex,PhysRevB.95.224424}. 
We call this double-$Q$ state a double-$Q$ chiral stripe state. 
 
To construct the ground-state phase diagram, we have performed the variational calculations in two steps.
First, we determine the optimal ordering vector $\bm{Q}_1$ by comparing the energies for each magnetic state described by Eq.~(\ref{1Q}). 
Second, we compare the energies between the optimal state obtained in the first step and double-$Q$ chiral stripe in Eq.~(\ref{2Q}). 
In the following, we present the results for $N_Q=42$, and the systems with $N_s=168^2$ unit cells. 
These results are qualitatively similar to the results obtained with other system sizes, e.g., $N_Q=24$ and $N_s=288^2$.

 \begin{figure}[t!]
 \begin{center}
\includegraphics[width=8.5cm]{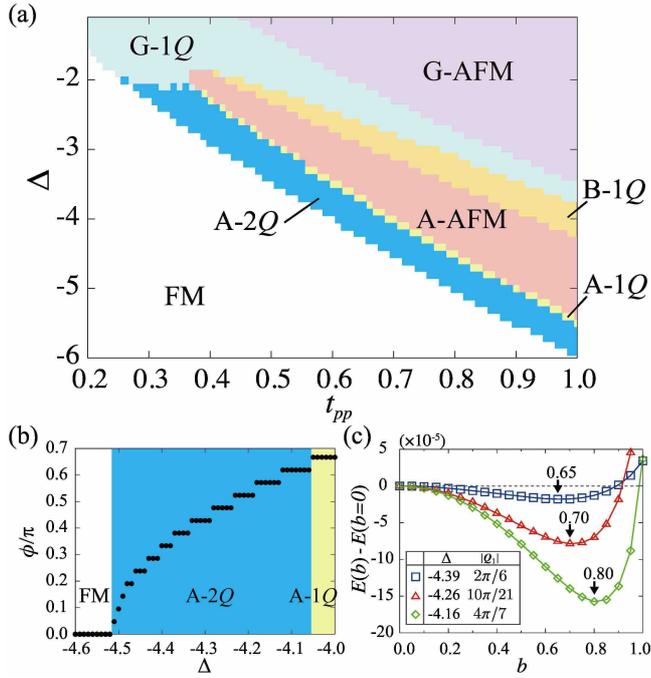}
\end{center}
\caption{(Color online).
(a) Ground-state phase diagram obtained by variational calculations as a function of $\Delta$ and $t_{pp}$. 
A(B, G)-1$Q$, A(G)-AFM and A-2$Q$ represent the A(B, G)-type single-$Q$ spiral, A(G)-type staggered AFM, and A-type double-$Q$ chiral stripe states, respectively.
 (b) $\Delta$ dependence of spiral pitch $\phi$ at $t_{pp}=0.7$. 
 (c)  $b$ dependence of internal energy of the A-type double-$Q$ chiral stripe measured from the single-$Q$ spiral at $t_{pp}=0.7$ for several $\Delta$.
 The arrows indicate the optimal $b$.
   }
\label{fig2}
\end{figure}

 Figure~\ref{fig2}(a) shows the ground-state phase diagram as a function of $\Delta$ and $t_{pp}$.
In the region for large negative $\Delta$, the FM state is stabilized by the double-exchange mechanism. 
As $\Delta$ is increased, the FM state is replaced by the 1$Q$ spiral and 2$Q$ chiral stripe states depending on $t_{pp}$. 
For small $0.2 \lesssim t_{pp} \lesssim 0.28$, the G-type 1$Q$ spiral (G-1$Q$) appears, while the A-type 2$Q$ chiral stripe (A-2$Q$) is realized for $0.28\lesssim t_{pp}$.
In the region where the A-2$Q$ state appears in the phase diagram, various magnetic phases are also realized with a further increase in $\Delta$: 
the A-2$Q$ state changes into the A-type 1$Q$ spiral (A-1$Q$), A-type staggered AFM (A-AFM), B-type 1$Q$ spiral (B-1$Q$), G-1$Q$, and G-type staggered AFM (G-AFM) states by increasing $\Delta$.

We show $\Delta$ dependence of the spiral angle $\phi$ at $t_{pp}=0.7$ for $-4.6 \leq \Delta \leq -4.0$ in Fig.~\ref{fig2}(b).
The optimal spiral pitch $\phi$ becomes nonzero at the phase boundary between the FM and A-2$Q$ states at $\Delta\sim-4.51$. 
 With increasing $\Delta$, the optimal pitch continuously increases from zero.
 Meanwhile, the optimal $b$($=b_{\rm opt}$) discontinuously increases from zero, where $b_{\rm opt}$ is between 0.45 and 0.85, as discussed below. 
Note that finite jumps of $\phi$ are attributed to the finite-size effect.
With further increases in $\Delta$, the A-2$Q$ state changes into the A-1$Q$ state with a continuity in $\phi$ and discontinuity in $b$. 
Similar behavior is obtained by varying $t_{pp}$ for a fixed $\Delta$.
 
To examine the value of $b_{\rm opt}$ in the A-2$Q$ state, we plot $b$ dependence of the internal energy per unit cell measured from that in the A-1$Q$ state ($b=0$) at $t_{pp}=0.7$ for $\Delta=-4.39$, $-4.26$, and $-4.16$ in Fig.~\ref{fig2}(c). 
The corresponding optimal ordering vectors are also shown in Fig.~\ref{fig2}(c).
There are two characteristic points against $b$:
a) $b_{\rm opt}$ is smaller than $1$, which means that there is a local minimum that exists as a function of $b$ for $0<b <1$ 
and b) the negative slope of the energy difference at $b=0$.
This tendency is similar to that in the Kondo lattice model~\cite{ozawa2016vortex}, which indicates that the perturbation analysis might be applicable to the present model. 
These two characteristics are common for other parameters in the the A-2$Q$ phase except for the region near the phase boundary between the A-2$Q$ and A-1$Q$ states for $0.4 \lesssim t_{pp}\lesssim 0.5$ where 
the slope of the energy at $b=0$ is positive.

\begin{figure}[t!]
\begin{center}
\includegraphics[width=8.5cm]{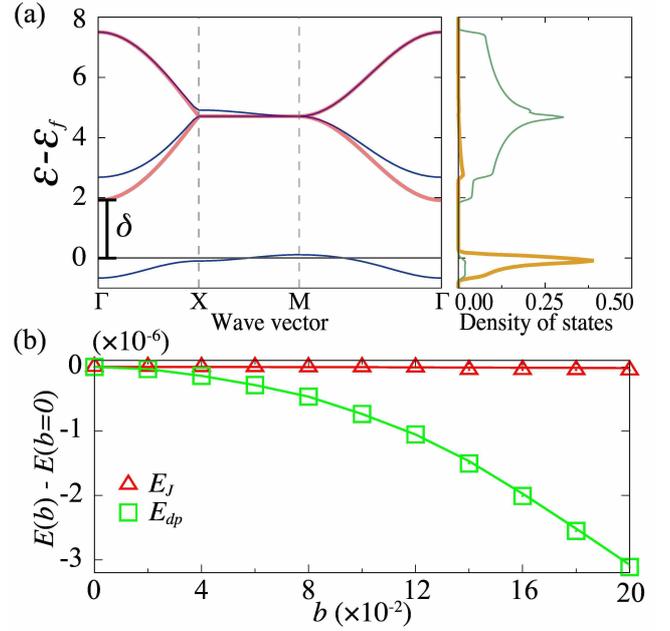}
\end{center}
\caption{(Color online).
(a) The left panel represents the energy band dispersion measured from the Fermi energy $\varepsilon_f$ (horizontal line) in the FM state at $t_{pp}=0.7$ and $\Delta=-4.6$.
The red thick and blue thin solid lines represent the $d$-$p$ hybridized spin-up and spin-down polarized bands, respectively.
$\delta$ is the energy difference between the energy at the bottom of the lowest unoccupied spin-up band and the Fermi energy.
$\Gamma$, X, and M stand for $(k_x, k_y)=(0, 0)$, $(\pi, 0)$, and $(\pi, \pi)$, respectively.
The right panel is the density of states for the $d$ (orange thick line)  and $p$ orbitals (green thin line).
  (b) The energy for spin-charge coupling terms ($E_J$) and otherwise ($E_{dp}$) in the Hamiltonian in Eq.~(\ref{H}) at $t_{pp}=0.7$, $\Delta=-4.39$ and $\bm{Q}_1=(2\pi/6, 0)$.
Each energy is measured from the energy in the single-$Q$ spiral. 
 }
\label{fig3}
\end{figure}    

The critical charge transfer energy $\Delta_c$ between the FM and A-2$Q$ states can be roughly estimated from the electronic band structure in the FM state~\cite{mostovoy2005helicoidal, mostovoy2005helicoidal_2}.
The left panel of Fig.~\ref{fig3}(a) shows the band structure in the FM state, where each eigenvalue $\varepsilon$ is measured from the Fermi energy $\varepsilon_f$ at $t_{pp}=0.7$ and $\Delta=-4.6$. 
There are three spin-up polarized bands (red thick solid lines) and three spin-down polarized bands (blue thin solid lines). 
The Fermi energy (horizontal line)  is located at the lowest spin-down polarized band. 
Note that one of the spin-up polarized bands is located at $\varepsilon \sim -J$ and fully occupied owing to the strong spin-charge coupling [see also Fig.~\ref{fig1}(b)].
The right panel of Fig.~\ref{fig3}(a) represents the density of states for the $d$ (orange thick line) and $p$ (green thin line) orbitals.
The upper four bands are attributed primarily to the $p$ orbital, while the lower two bands are attributed to the $d$ orbital.

The instability toward the single-$Q$ spiral or double-$Q$ chiral stripe states from the FM state occurs when the lowest unoccupied spin-up $p$ band is close enough to the highest occupied spin-down $d$
 band~\cite{mostovoy2005helicoidal, mostovoy2005helicoidal_2}. 
In that situation, the hybridization between the spin-up $p$ band and spin-down $d$ band lowers the energies of the occupied states, although the energy gain with respect to the double-exchange mechanism becomes smaller owing to the relative angle of the local spins through the hybridization.
The energy difference $\delta$ between the energy at the bottom of the lowest unoccupied spin-up band $\varepsilon_{\uparrow}^{\Gamma}$ and the Fermi energy $\varepsilon_f$ is given by
\begin{align}
\delta &\approx \varepsilon_{\uparrow}^{\Gamma}-\left\{ \varepsilon_{\downarrow}^{\Gamma}+\frac{5}{6} \left(\varepsilon_{\downarrow}^{\text{\rm M}}-\varepsilon_{\downarrow}^{\Gamma} \right)\right\} \nonumber \\
&\approx \frac{1}{12} \left( -11\Delta -44t_{pp}+\sqrt{(\Delta+4t_{pp})^2+32t_{pd}^2}\  \right),
\end{align}
where we set $\varepsilon_f\approx  \varepsilon_{\downarrow}^{\Gamma}+\frac{5}{6} \left(\varepsilon_{\downarrow}^{\text{\rm M}}-\varepsilon_{\downarrow}^{\Gamma} \right)$ by assuming that the density of states of the lowest occupied spin-down band is constant.
In the second line, we use $J \gg |\Delta|,  t_{pp}, t_{pd}$.
The critical charge transfer energy $\Delta_c$ is given by the condition to satisfy $\delta=0$, which  is represented by
\begin{align}
\Delta_c=-4t_{pp} + \frac{2}{\sqrt{15}}t_{pd}.
\label{delta}
\end{align}
Thus, $\Delta_c$ is proportional to $t_{pp}$, which is qualitatively consistent with the phase boundary in Fig.~\ref{fig1}(c).
Moreover, Eq.~(\ref{delta}) implies that the phase transition does not depend on the spin-charge coupling $J$, which is in contrast to the result in the weak coupling regime, where $J$ is essential to inducing the instability toward the single-$Q$ spiral and/or double-$Q$ chiral stripe states~\cite{ozawa2016vortex, PhysRevB.95.224424}. 

To clearly show that the spin-charge coupling is less important in stabilization of the double-$Q$ chiral stripe, we compare the energetic contributions from the spin-charge coupling term and otherwise in the Hamiltonian in Eq.~(\ref{H}). 
The spin-charge coupling energy is given by $E_J=-J\langle \sum_{i,\sigma,\sigma'} d^{\dagger}_{i\sigma} \bm{\sigma}_{\sigma\sigma'}d_{i\sigma'}  \cdot \bm{S}_i-\sum_{i,c,\sigma} p^{\dagger}_{i+\frac{c}{2}\sigma}p_{i+\frac{c}{2}\sigma}\rangle$, and the energy from other contributions is given by $E_{dp}=\langle t_{pd}\sum_{i,c,\sigma} (d^{\dagger}_{i\sigma}P_{ic,\sigma}+P^{\dagger}_{ic,\sigma}d_{i\sigma})-t_{pp}\sum_{i,c\neq c',\sigma} P^{\dagger}_{ic,\sigma}P_{ic',\sigma}-\Delta\sum_{i,c,\sigma} p^{\dagger}_{i+\frac{c}{2}\sigma}p_{i+\frac{c}{2}\sigma}\rangle$.

Figure~\ref{fig3}(b) shows the $b$ dependence of $E_{J}$ and $E_{dp}$ at $t_{pp}=0.7$ and $\Delta=-4.39$ in the small $b$ region.
The result reveals that energy gain in the double-$Q$ chiral stripe is due to the $d$-$p$ energy $E_{dp}$.  
On the other hand, there is almost no energy gain from the spin-charge coupling energy $E_J$. 
In other words, the effect of $J$ appears through the kinetic motion of itinerant electrons rather than the direct exchange coupling.
This result indicates that other effective spin-spin interactions emerge through the $d$-$p$ hybridization, which gives rise to the double-$Q$ chiral stripe.
Such a study to obtain effective interactions in the strong spin-charge coupling regime remains an interesting problem for future study.

Finally, we discuss the stability of the double-$Q$ chiral stripe by taking into account other interactions.
We here introduce the AFM superexchange interaction between the localized spins, 
$\mathcal{H}_{\rm{SE}}=J_{\rm{SE}}\sum_{i,c=x,y} \bm{S}_i \cdot \bm{S}_{i+c}$ for $J_{\rm{SE}}>0$, which sometimes leads to short-period multiple-$Q$ structures even in the double-exchange model~\cite{PhysRevLett.105.216405,PhysRevB.62.13816}.

\begin{figure}[t]
\begin{center}
\includegraphics[width=8.0cm]{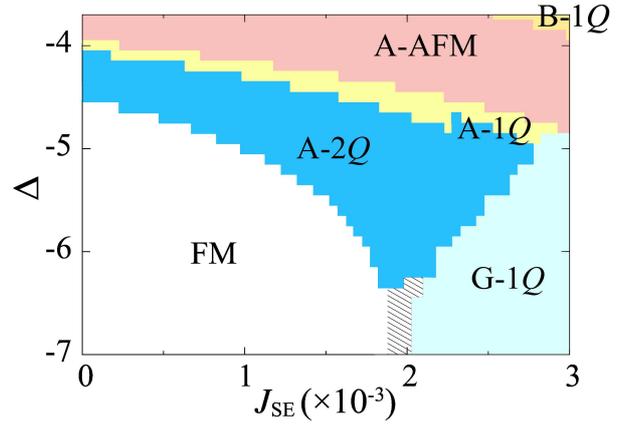}
\end{center}
\caption{(Color online). Ground-state phase diagram as a function of $\Delta$ and $J_{\rm{SE}}$ at $t_{pp}=0.7$. 
In the hatched area, it is difficult to determine the ground state because of the finite-size effect.} 
\label{fig4}
\end{figure}

Figure~\ref{fig4} shows the ground-state phase diagram in the $J_{\rm SE}$-$\Delta$ plane at $t_{pp}=0.7$.
By introducing $J_{\rm SE}$, the A-2$Q$ state becomes more stable.
The energy change by $J_{\rm SE}$ in the double-$Q$ chiral stripe state is evaluated by expanding the square root in Eq.~(\ref{2Q}) with respect to $b^2$:
$\sqrt{1-b^2+b^2\cos^2\bm{Q}_2\cdot\bm{r}_i} = \sum_{m} C_{2m} \cos 2m \bm{Q}_2\cdot \bm{r}_i$ $(m=0,1,2,\cdots)$, where coefficients are given by $C_0=1-\frac{1}{4}b^2-\frac{3}{64}b^4$, $C_2=\frac{1}{4}b^2+\frac{1}{16}b^4$, and $C_4=-\frac{1}{64}b^4$ up to $b^4$.   
In the end, the superexchange energy $E^{2Q}_{\rm SE}$ in the double-$Q$ chiral stripe is expressed as 
\begin{align}
E^{2Q}_{\rm SE}&=J_{\rm{SE}}\sum_{c=x,y} \left(\cos Q_{1c} - \frac{b^4}{16}\cos Q_{1c} \sin^2 Q_{2c}\right) +\mathcal{O}(b^6)\nonumber \\
&\approx E^{1Q}_{\rm SE} - J_{\rm{SE}}\frac{b^4}{16}\sum_{c=x,y} \cos Q_{1c} \sin^2 Q_{2c},
\end{align}
where $Q_{1c}$($Q_{2c}$) is the $c$-component of $\bm{Q}_1$($\bm{Q}_2$), and $E^{1Q}_{\rm SE}=J_{\rm{SE}}\sum_c \cos Q_{1c}$ is the superexchange energy in the single-$Q$ spiral state. 
Specifically, the energy gain by $J_{\rm SE}$ in the A-2$Q$ state compared to the A-1$Q$ state is given by $J_{\rm{SE}}\frac{b^4}{16}\sin^2\phi$.
Thus, the AFM superexchange interaction favors the A-2$Q$ state compared to the FM and A-1$Q$ states~\cite{comment}.
In the region for large $J_{\rm SE}$, the A-1$Q$ and A-2$Q$ states are replaced by the G-1$Q$ state, as $J_{\rm SE}$ favors the G-type magnetic orders.
 
In summary, by using the variational calculations, we have clarified that the $d$-$p$ hybridization provides another route to stabilize double-$Q$ chiral stripe states in the ground state for the strong spin-charge coupling.  
We also found that the antiferromagnetic superexchange interaction favors the double-$Q$ chiral stripe state. 
Our results will be helpful in attempts to unveil the microscopic origin of a plethora of multiple-$Q$ states in transition metal oxides with the strong spin-charge (Hund's-rule) coupling.
The itinerant magnet SrFeO$_3$ is one of the candidate materials, as the observed double-$Q$ state at low temperatures is similar to the double-$Q$ chiral stripe state in the present study~\cite{ishiwata2018emergent}, although the detailed lattice structures are different from each other.
A detailed comparison will be left for future study. 

\begin{acknowledgment}
S. H. is grateful to S. Ishiwata for fruitful discussions. 
This research was supported by JSPS KAKENHI Grants Number JP18K13488. 
This work was also supported by the Toyota Riken Scholarship. 
Parts of the computation in this work were carried out at the Supercomputer Center, Institute for Solid State Physics, the University of Tokyo.
We would like to thank Editage (www.editage.com) for English language editing.
\end{acknowledgment}

\bibliographystyle{jpsj}
\bibliography{17630}

\begin{thebibliography}{10}

\bibitem{PhysRevB.45.13544}
D.~Loss and P.~M. Goldbart, Phys. Rev. B {\bfseries 45},  13544 (1992).

\bibitem{PhysRevLett.83.3737}
J.~Ye, Y.~B. Kim, A.~J. Millis, B.~I. Shraiman, P.~Majumdar, and Z.~Te\ifmmode
  \check{s}\else \v{s}\fi{}anovi\ifmmode~\acute{c}\else \'{c}\fi{}, Phys. Rev.
  Lett. {\bfseries 83},  3737 (1999).

\bibitem{PhysRevB.62.R6065}
K.~Ohgushi, S.~Murakami, and N.~Nagaosa, Phys. Rev. B {\bfseries 62},  R6065
  (2000).

\bibitem{PhysRevLett.87.116801}
R.~Shindou and N.~Nagaosa, Phys. Rev. Lett. {\bfseries 87},  116801 (2001).

\bibitem{taguchi2001spin}
Y.~Taguchi, Y.~Oohara, H.~Yoshizawa, N.~Nagaosa, and Y.~Tokura, Science
  {\bfseries 291},  2573 (2001).

\bibitem{PhysRevLett.101.156402}
I.~Martin and C.~D. Batista, Phys. Rev. Lett. {\bfseries 101},  156402 (2008).

\bibitem{PhysRevLett.105.266405}
Y.~Kato, I.~Martin, and C.~D. Batista, Phys. Rev. Lett. {\bfseries 105},
  266405 (2010).

\bibitem{batista2016frustration}
C.~D. Batista, S.-Z. Lin, S.~Hayami, and Y.~Kamiya, Rep. Prog. Phys. {\bfseries
  79},  084504 (2016).

\bibitem{DZYALOSHINSKY1958241}
I.~Dzyaloshinsky, J. Phys. Chem. Solids {\bfseries 4},  241  (1958).

\bibitem{PhysRev.120.91}
T.~Moriya, Phys. Rev. {\bfseries 120},  91 (1960).

\bibitem{roessler2006spontaneous}
U.~K. R\"{o}{\ss}ler, A.~Bogdanov, and C.~Pfleiderer, Nature {\bfseries 442},
  797 (2006).

\bibitem{PhysRevLett.96.207202}
B.~Binz, A.~Vishwanath, and V.~Aji, Phys. Rev. Lett. {\bfseries 96},  207202
  (2006).

\bibitem{PhysRevB.80.054416}
S.~D. Yi, S.~Onoda, N.~Nagaosa, and J.~H. Han, Phys. Rev. B {\bfseries 80},
  054416 (2009).

\bibitem{PhysRevB.92.134405}
R.~Keesman, A.~O. Leonov, P.~van Dieten, S.~Buhrandt, G.~T. Barkema, L.~Fritz,
  and R.~A. Duine, Phys. Rev. B {\bfseries 92},  134405 (2015).

\bibitem{PhysRevLett.108.017206}
T.~Okubo, S.~Chung, and H.~Kawamura, Phys. Rev. Lett. {\bfseries 108},  017206
  (2012).

\bibitem{leonov2015multiply}
A.~Leonov and M.~Mostovoy, Nat. Commun. {\bfseries 6},  8275 (2015).

\bibitem{PhysRevB.93.064430}
S.-Z. Lin and S.~Hayami, Phys. Rev. B {\bfseries 93},  064430 (2016).

\bibitem{PhysRevB.93.184413}
S.~Hayami, S.-Z. Lin, and C.~D. Batista, Phys. Rev. B {\bfseries 93},  184413
  (2016).

\bibitem{PhysRevLett.108.096401}
Y.~Akagi, M.~Udagawa, and Y.~Motome, Phys. Rev. Lett. {\bfseries 108},  096401
  (2012).

\bibitem{ozawa2016vortex}
R.~Ozawa, S.~Hayami, K.~Barros, G.-W. Chern, Y.~Motome, and C.~D. Batista, J.
  Phys. Soc. Jpn. {\bfseries 85},  103703 (2016).

\bibitem{PhysRevB.95.224424}
S.~Hayami, R.~Ozawa, and Y.~Motome, Phys. Rev. B {\bfseries 95},  224424
  (2017).

\bibitem{PhysRev.96.99}
M.~A. Ruderman and C.~Kittel, Phys. Rev. {\bfseries 96},  99 (1954).

\bibitem{10.1143/PTP.16.45}
T.~Kasuya, Prog. Theor. Phys. {\bfseries 16},  45 (1956).

\bibitem{PhysRev.106.893}
K.~Yosida, Phys. Rev. {\bfseries 106},  893 (1957).

\bibitem{akagi2010spin}
Y.~Akagi and Y.~Motome, J. Phys. Soc. Jpn. {\bfseries 79},  083711 (2010).

\bibitem{barros2013efficient}
K.~Barros and Y.~Kato, Phys. Rev. B {\bfseries 88},  235101 (2013).

\bibitem{PhysRevB.90.060402}
S.~Hayami and Y.~Motome, Phys. Rev. B {\bfseries 90},  060402 (2014).

\bibitem{PhysRevLett.118.147205}
R.~Ozawa, S.~Hayami, and Y.~Motome, Phys. Rev. Lett. {\bfseries 118},  147205
  (2017).

\bibitem{PhysRevB.99.094420}
S.~Hayami and Y.~Motome, Phys. Rev. B {\bfseries 99},  094420 (2019).

\bibitem{PhysRevLett.120.097201}
G.~G. Marcus, D.-J. Kim, J.~A. Tutmaher, J.~A. Rodriguez-Rivera, J.~O. Birk,
  C.~Niedermeyer, H.~Lee, Z.~Fisk, and C.~L. Broholm, Phys. Rev. Lett.
  {\bfseries 120},  097201 (2018).

\bibitem{takagi2018multiple}
R.~Takagi, J.~White, S.~Hayami, R.~Arita, D.~Honecker, H.~R{\o}nnow, Y.~Tokura,
  and S.~Seki, Sci. Adv. {\bfseries 4},  eaau3402 (2018).

\bibitem{kurumaji2019skyrmion}
T.~Kurumaji, T.~Nakajima, M.~Hirschberger, A.~Kikkawa, Y.~Yamasaki,
  H.~Sagayama, H.~Nakao, Y.~Taguchi, T.-h. Arima, and Y.~Tokura, Science
  {\bfseries 365},  914 (2019).

\bibitem{hirschberger2018skyrmion}
M.~Hirschberger, T.~Nakajima, S.~Gao, L.~Peng, A.~Kikkawa, T.~Kurumaji,
  M.~Kriener, Y.~Yamasaki, H.~Sagayama, H.~Nakao, K.~Ohishi, K.~Kakurai,
  Y.~Taguchi, X.~Yu, T.~Arima, and Y.~Tokura, arXiv:1812.02553 ,  (2018).

\bibitem{PhysRev.82.403}
C.~Zener, Phys. Rev. {\bfseries 82},  403 (1951).

\bibitem{PhysRev.100.675}
P.~W. Anderson and H.~Hasegawa, Phys. Rev. {\bfseries 100},  675 (1955).

\bibitem{PhysRev.118.141}
P.~G. de~Gennes, Phys. Rev. {\bfseries 118},  141 (1960).

\bibitem{mostovoy2005helicoidal}
M.~Mostovoy, Phys. Rev. Lett. {\bfseries 94},  137205 (2005).

\bibitem{mostovoy2005helicoidal_2}
M.~Mostovoy, J. Phys.: Condens. Matter {\bfseries 17},  S753 (2005).

\bibitem{PhysRevB.84.054427}
S.~Ishiwata, M.~Tokunaga, Y.~Kaneko, D.~Okuyama, Y.~Tokunaga, S.~Wakimoto,
  K.~Kakurai, T.~Arima, Y.~Taguchi, and Y.~Tokura, Phys. Rev. B {\bfseries 84},
   054427 (2011).

\bibitem{ishiwata2018emergent}
S.~Ishiwata, T.~Nakajima, J.-H. Kim, D.~S. Inosov, N.~Kanazawa, J.~S. White,
  J.~L. Gavilano, R.~Georgii, K.~Seemann, G.~Brandl, P.~Manuel, D.~D.
  khalyavin, S.~Seki, Y.~Tokunaga, M.~Kinoshita, Y.~W. Long, Y.~Kaneko,
  Y.~Taguchi, T.~Arima, B.~Keimer, and Y.~Tokura, arXiv:1806.02309 ,  (2018).

\bibitem{PhysRevMaterials.3.084404}
P.~C. Rogge, R.~J. Green, R.~Sutarto, and S.~J. May, Phys. Rev. Mater.
  {\bfseries 3},  084404 (2019).

\bibitem{comment3}
We confirmed that the double-Q chiral stripe state presented in the following
  is also stabilized when we consider an itinerant $d_{x^2-y^2}$ orbital
  instead of $d_{3z^2-r^2}$.

\bibitem{comment2}
For example, the continuous phase transition between the FM state and the
  single-$Q$ spiral state is found when the Fermi level lies at the lowest
  spin-down band in the FM state. See also Fig. 3(a).

\bibitem{PhysRevLett.105.216405}
S.~Kumar and J.~van~den Brink, Phys. Rev. Lett. {\bfseries 105},  216405
  (2010).

\bibitem{PhysRevB.62.13816}
D.~F. Agterberg and S.~Yunoki, Phys. Rev. B {\bfseries 62},  13816 (2000).

\bibitem{comment}
The energy gain in the G(B)-2$Q$ state compared to the G(B)-1$Q$ state is given
  by $J_{\rm{SE}}\frac{b^4}{8}\cos\phi\sin^2\phi$
  $(-J_{\rm{SE}}\frac{b^4}{16}\sin^2\phi)$, which means the G-2$Q$ for
  $\phi<\frac{\pi}{2}$ state is stabilized by $J_{\rm{SE}}$, while the G-2$Q$
  state $\phi>\frac{\pi}{2}$ and the B-2$Q$ state are destabilized.

\end{thebibliography}

\end{document}